# Energy loss of the electron system in individual single-walled carbon nanotubes

Daniel F. Santavicca,[1] Joel D. Chudow,[1] Daniel E. Prober,[1*] Meninder S. Purewal,[2] and Philip Kim[2]

[1] *Department of Applied Physics, Yale University*
[2] *Departments of Physics and Applied Physics, Columbia University*

[*]e-mail: daniel.prober@yale.edu

ABSTRACT: We characterize the energy loss of the non-equilibrium electron system in individual metallic single-walled carbon nanotubes at low temperature. Using Johnson noise thermometry, we demonstrate that, for a nanotube with ohmic contacts, the dc resistance at finite bias current directly reflects the average electron temperature. This enables a straightforward determination of the thermal conductance associated with cooling of the nanotube electron system. In analyzing the temperature- and length-dependence of the thermal conductance, we consider contributions from acoustic phonon emission, optical phonon emission, and hot electron outdiffusion.

Carbon nanotubes are attractive for a number of device applications because of their ability to support extremely large current densities, of order $10^9$ A/cm$^2$.[1] Such large current densities can lead to significant Joule heating, and hence self-heating effects are important in determining the performance limits of nanotube-based devices. These self-heating effects also provide a tool for studying the non-equilibrium electron properties of this unique one-dimensional conductor. Specifically, we use Joule heating to study the inelastic processes by which the nanotube electron system loses energy to the environment. We focus on a high-quality individual SWNT on an insulating substrate below room temperature. We analyze our results in terms of theoretical predictions for acoustic phonon emission, optical phonon emission, and hot electron outdiffusion.

Several previous works have studied heating effects in individual single-walled nanotubes (SWNT) at room temperature and above. Park *et al.* report current-voltage curves of a high quality individual SWNT on an insulating substrate at a bath (substrate) temperature $T_b = 300$ K.[2] At high bias, current saturation due to optical phonon emission is observed, consistent with previous studies.[1] At low bias, the inferred electron mean free path is consistent with the calculated mean-free path for electron-acoustic phonon scattering. Other works have studied the effects of Joule and optical heating on the nanotube lattice temperature, which can differ from the electron temperature. Pop *et al.* and Maune *et al.* estimate the phonon interface thermal conductance between an individual SWNT and an insulating substrate from measurements of electrical breakdown.[3,4] Hsu *et al.* use the shift of the G band Raman frequency to infer the local phonon temperature in a suspended SWNT heated above room temperature by laser illumination,[5] enabling a comparison of the contact and the internal thermal resistance. Shi *et al.* use a scanning thermal microscope to determine the local phonon temperature along the length of a Joule-heated individual SWNT on an insulating substrate at $T_b = 300$ K.[6] In the present work, we measure the average electron temperature of a Joule-heated nanotube. We focus on the low temperature regime, 4 K $< T <$ 200 K. Lower temperatures result in greater thermal decoupling of the electron and phonon systems, facilitating the study of low energy inelastic processes of the electron system. We compare our results to extrapolations of the previous higher temperature measurements.

The nanotube studied in this work was grown using chemical vapor deposition on a degenerately doped silicon (Si) substrate with a 500 nm thick oxide (SiO$_2$). The growth procedure produces nanotubes that are up to millimeters in length.[7] Palladium electrodes are then deposited at various separations along an individual nanotube. The Si substrate is used as a global back gate. We report two-terminal electrical measurements of nanotube segment lengths of 2, 5, 20 and 50 μm, all of which are separately-contacted sections of the same nanotube. The diameter of this nanotube is 2.0 +/- 0.2 nm, measured with an atomic force microscope,[8] and the saturation current was measured to be < 30 μA, ensuring that it is an individual single-walled tube.[1] It is a small band gap (<100 meV) quasi-metallic nanotube of unknown chirality. This is the same nanotube referred to as sample M1 in Ref. [8]. The properties reported in that previous work were found to be the same in the present work. All measurements were conducted at a back gate voltage of -30 V, where the two-terminal conductance is a maximum and is insensitive to small variations in the gate potential.

The dc resistance, $R_{dc} = V_{dc}/I_{dc}$, is shown in Figure 1 as a function of temperature for the 5 μm long nanotube section measured with a small bias current ($I_{dc}$ = 0.3 μA). Measurements of all four nanotube lengths indicate a contact resistance $R_c \approx$ 8 kΩ, close to the quantum-limited two-terminal contact resistance of $R_Q/4 = h/4e^2 \approx$ 6.4 kΩ for a one dimensional channel with four subbands, and an internal resistance $R_{int} \approx$ 1 kΩ/μm at 4 K that increases to ≈ 12 kΩ/μm at 300 K; these same results were also seen in Ref. [8]. The approximately linear temperature dependence of the dc resistance observed above 50 K is consistent with electron-acoustic phonon inelastic scattering.[2,9]

In Figure 1 (inset) we plot the measured dc resistance of the 5 μm nanotube as a function of $I_{dc}$ at bath temperatures $T_b$ = 4.2 K and 77 K. For measurements at $T_b$ < 20 K, a local maximum in $R_{dc}$ is seen at zero bias current. This zero-bias anomaly (ZBA) has been discussed in a number of previous works, and has alternatively been attributed to a reduced density of states for tunneling into a Luttinger liquid[1,10] or to Coulomb blockade.[11,12] In either case, the ZBA is related to non-ohmic contacts. At $T_b$ = 4.2 K, as the bias current is increased above 0.5 μA, the contacts recover ohmic behavior and $R_{dc}$ displays a monotonic increase with increasing bias current. We measure up to $I_{dc}$ = 5 μA, which is large enough to show significant heating effects but still well below the saturation current.

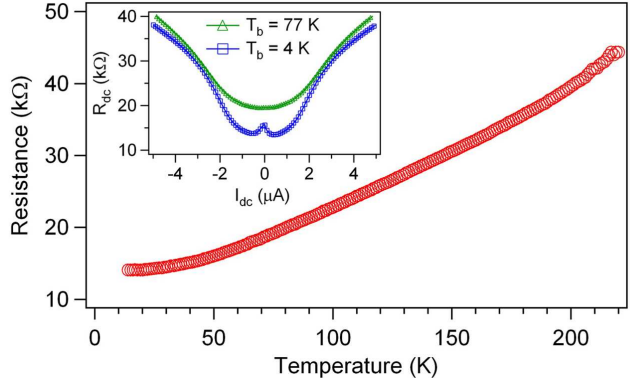

FIGURE 1. DC resistance of 5 µm long nanotube as a function of temperature measured with a dc bias current of 0.3 µA. Inset: DC resistance as a function of dc bias current at bath temperatures of 4.2 K and 77 K.

The increase in $R_{dc}$ with increasing $I_{dc}$ is due to Joule heating of the electron system in the nanotube. We use Johnson noise to directly determine the average electron temperature as a function of $I_{dc}$. This allows us to establish that the dc resistance is a measure of the average electron temperature. Thus, the $R_{dc}(T_b)$ data can be used to assign a temperature to the electron system in the $R_{dc}(I_{dc})$ data. This should be useful for other researchers, as the $R_{dc}(I_{dc})$ data are much easier to collect than the Johnson noise data.

The understanding of the Johnson noise measurements is as follows. For a resistor with a uniform electron temperature $T_e$, the Johnson noise power coupled into a matched load is, in the low frequency limit ($hf \ll k_B T_e$), $P_J = k_B T_e B$, where $B$ is the measurement bandwidth. The quantity $P_J/k_B B$ is the Johnson noise temperature $T_J$, and for a spatially uniform electron temperature $T_J = T_e$. For a temperature-dependent resistor with a spatial temperature distribution (and no contact resistance), $T_J = \int_0^L T_e(x) r(x) dx \Big/ R_{tot}$, where $T_e(x)$ and $r(x)$ are the position-dependent electron temperature and the resistance per unit length, respectively, and $R_{tot}$ is the total resistance.

To find the average electron temperature, we need to model the electron temperature profile $T_e(x)$ within a Joule-heated nanotube. To do this, we use the one-dimensional steady-state heat flow equation,

$$\frac{\partial}{\partial x}\left(g_{diff}\frac{\partial T_e}{\partial x}\right) + p_{NT} - p_{ph} = 0. \tag{1}$$

The first term is due to hot electron diffusion, with $x$ the position along the length of the nanotube, $0 \leq x \leq L$. This model is valid provided that $L$ is greater than the inelastic electron-electron scattering length, which implies a well-defined local electron temperature $T_e(x)$. Previous measurements of the nanotube electron energy distribution via tunneling spectroscopy found that a non-thermal electron distribution is only exhibited at temperatures well below 4 K for $L = 1 - 2$ μm.[13] Hence, the assumption of a well-defined local electron temperature should be valid for all samples over the entire temperature range studied in the present work.

The electron diffusion thermal conductance per unit length $g_{diff}(x)$ is determined from the Wiedemann-Franz law, $g_{diff}(x) = \mathcal{L} T_e(x)/r(x)$, with $\mathcal{L}$ the Lorenz number. We approximate the internal resistance per unit length $r(x) = \alpha T_e(x)/L$, with $\alpha$ the slope of a linear fit to the $R_{dc}(T_b)$ data in Figure 1 above 50 K. The Joule power per unit length dissipated internal to the nanotube is $p_{NT} = I_{dc}^2 r(x)$. We assume that the power dissipated by the contact resistance remains in the relatively massive contacts, which act as thermal reservoirs. The total contact resistance (for the two contacts in series) is $R_c$, so the total two-terminal resistance is $R = R_c + \int_0^L r(x)dx$. $p_{ph}(T_e)$ is the power removed from the nanotube by phonon emission per unit length. The thermal conductance $G_c$ across the contact at $x = 0$ is incorporated via the boundary condition

$$g_{diff}\frac{\partial T_e}{\partial x}\bigg|_{x=0} = G_c\left[T_e(x=0) - T_b\right]. \tag{2}$$

This is determined from the Wiedemann-Franz law, $G_c = \mathcal{L} T_{avg}/(R_c/2)$, with $(R_c/2) \approx 4$ kΩ the electrical resistance at each contact. A similar expression is used for the contact at $x = L$. For electron thermal transport across the contacts, we assume that the temperature in the Wiedemann-Franz law for $G_c$ is the average of the temperatures on each side of the barrier, $T_{avg}$, as is the case for a quantum point contact.[14]

To analyze the Johnson noise data, we solve eqs 1 and 2 for $T_e(x)$. $p_{ph}(T_e)$ is determined from measurements of the longest nanotube segment, for which end effects are small (discussed later). For simplicity, we assume that $p_{NT}$ is spatially uniform. We find that the electron temperature variation along the length of the nanotube is always smaller than the temperature change across the contacts. Thus, $r(x)$ is relatively constant, so $p_{NT}(x) = I_{dc}^2 r(x)$ is also fairly uniform. As an example, we show in Figure 2 (inset) the calculated electron temperature profile for the 5 μm nanotube sample at $T_b = 77$ K with $I_{dc} = 1.5$ μA and 2 μA. For $I_{dc} = 2$ μA, the calculated temperature profile corresponds to an average electron temperature $T_e = 118.2$ K. We can also use the calculated $T_e(x)$ to determine $r(x)$, and then calculate $T_J$ using the formula above. This yields $T_J = 118.5$ K. Hence, we can take the measured Johnson noise temperature to be equal to the average electron temperature to a good approximation.

To measure the Johnson noise, we employ a differential measurement technique with a bias current that switches between zero current and finite current at low frequency. The noise is measured with a 50 Ω microwave amplifier through a bandpass filter with a 10 MHz bandwidth centered at approximately 50 MHz. This avoids $1/f$ noise, which has been shown to be significant in carbon nanotubes, and to depend on the bias current.[15] We account for the coupling mismatch between the nanotube and the 50 Ω amplifier at each value of bias current. The amplifier output is coupled to a diode to measure power, and the diode response is read on a lock-in amplifier that is ac synchronized to the on-off bias current. As before, we assume that only $R_{int}$ is heated by the bias current and that $R_c$ is temperature-independent. We have also calculated the change in noise due to intrinsic thermodynamic fluctuations, which result in resistance fluctuations due to the nanotube's temperature-dependent resistance.[16,17] We found that this is more than an order of magnitude smaller than the change in Johnson noise, and hence is neglected in the present analysis.

We compare in Figure 2 the temperature increase of the 5 μm nanotube sample determined from the Johnson noise measurement and from using the thermal equilibrium $R_{dc}(T_b)$ data (for $I_{dc} \approx 0$ ) to assign a temperature to the non-equilibrium electron system from the $R_{dc}(I_{dc})$ data. A bath temperature of 77 K is used to avoid the ZBA feature. The standard deviation of the measured Johnson noise temperature is approximately the same size as the data points. We see reasonable agreement between the temperatures determined using these two different approaches. We conclude that, away from the ZBA feature, the dc resistance is a measure of the average electron temperature for both the equilibrium ($I_{dc} \approx 0$) and the non-equilibrium (large $I_{dc}$) cases. We note that this technique of using the electrical resistance as a thermometer of the electron system has been studied previously in thin metal films[18] and is the operating principle of the bolometric detector.[16,17]

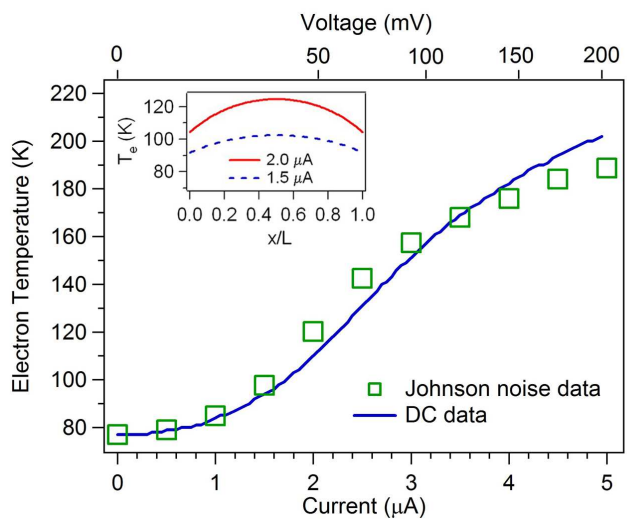

FIGURE 2. Average electron temperature $T_e$ of the 5 µm nanotube as a function of bias current at $T_b = 77$ K determined from Johnson noise thermometry (squares) and from the dc data in Figure 1 (solid line). The corresponding bias voltage is shown on the top axis. Inset: calculated electron temperature profile for 5 µm nanotube at $T_b = 77$ K with $I_{dc} = $ 1.5 µA and 2 µA.

As seen in Figure 2, the average electron temperature $T_e$ can be determined from $R_{dc}(I_{dc})$. We can then use this to determine the thermal conductance for cooling of the electron system. The total Joule power dissipated internal to the nanotube is $P_{NT} = I_{dc}^2 R_{int}$. The thermal conductance for cooling of the electron system in the nanotube is $G = dP_{NT}/dT_e$. This is plotted as a function of $T_e$ in Figure 3 for all four nanotube lengths. A smoothing function is applied to the data to minimize the noise from the numerical differentiation. Although we measure at $T_b = 4.2$ K, we only present $G$ for $T_e > 20$ K because of the ZBA feature at lower temperature.

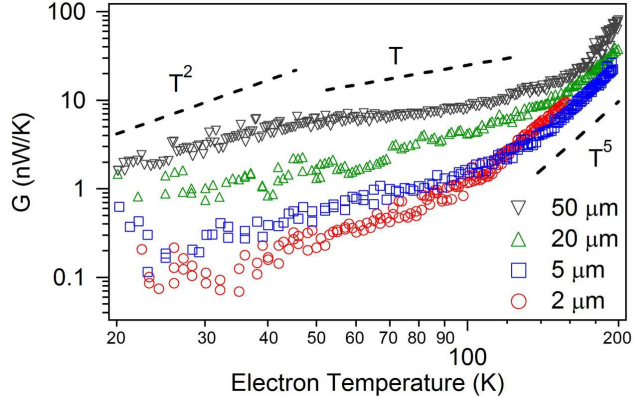

FIGURE 3. Thermal conductance as a function of average electron temperature for 2, 5, 20 and 50 µm nanotube lengths at a bath temperature of 4.2 K. Dashed lines illustrate $T$, $T^2$ and $T^5$ dependencies.

At temperatures above 120 K, $G$ increases rapidly with increasing temperature, approximately as $T_e^5$. We discuss that regime later, and focus here on the behavior for $T_e$ < 120 K. In this regime, we first consider the length-dependence of the measured thermal conductance. For sufficiently short nanotubes, the dominant cooling path will be the outdiffusion of electrons at temperature $T_e > T_b$ into the contacts. For long nanotubes, the dominant cooling path will instead be into the substrate via the emission of acoustic phonons. These parallel cooling paths are shown schematically in the inset of Figure 4a. In this simplified model, the electron system is represented by a single temperature $T_e$ and the contacts and substrate are thermal reservoirs at a temperature $T_b$.

We plot in Figure 4a the measured thermal conductance as a function of length at $T_e = 80$ K and $T_b = 4$ K. The data are fit to the sum of the two parallel thermal conductances seen in the inset schematic, $G = G_{sub} + 2G_c$. In calculating the electron temperature profile (discussed previously), we found that cooling via electron outdiffusion is limited largely by the contact thermal conductance $G_c$ rather than by the nanotube's internal thermal conductance for electron diffusion. Hence the total conductance for cooling via outdiffusion is simplified as $2G_c$, where the factor of 2 comes from considering diffusion out both ends of the nanotube. $G_c$ is calculated from the Wiedemann-Franz law for each contact, as before. For $T_e = 80$ K and $T_b = 4$ K, we determine from the fit in Figure 4a that $G_c \approx 0.25$ nW/K, in good agreement with the computed Wiedemann-Franz value of 0.26 nW/K. In Figure 4a, $G_{sub}$ is treated as a fitting parameter, with the requirement that it scales linearly with the nanotube length. We find $G_{sub} \approx (1.2 \times 10^{-4} L)$ in units of W/K, where $L$ is nanotube length in meters. The crossover length between diffusion-dominated and substrate-dominated cooling is $L \approx 3$ µm. This crossover length does not vary significantly with $T_e$ because both $G_{sub}$ and $G_c$ display an approximately linear dependence on $T_e$ (for $T_e >> T_b$).

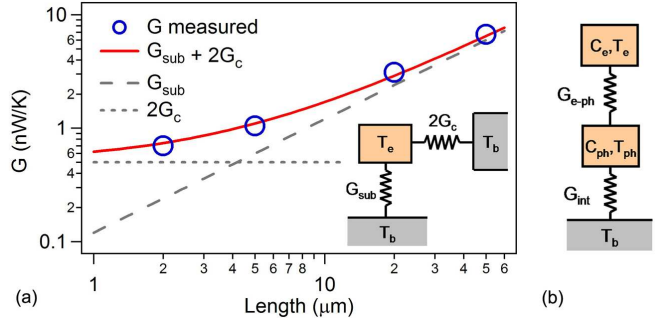

FIGURE 4. (a) Thermal conductance as a function of nanotube length at $T_e = 80$ K and $T_b = 4$ K. Solid line is a fit the thermal model illustrated in the inset. In this model, the electron system loses energy via two parallel thermal paths corresponding to cooling into the contacts via electron outdiffusion ($2G_c$) and cooling into the substrate via phonon emission ($G_{sub}$). (b) Thermal model for cooling of the nanotube electron system for long nanotube lengths and low bias currents. The phonon system in the nanotube is coupled to the electron system via the electron-acoustic phonon thermal conductance $G_{e-ph}$. The phonon system in the nanotube is coupled to the substrate via the phonon interface thermal conductance $G_{int}$.

Next we consider the temperature-dependence of the thermal conductance data, again for $T_e < 120$ K. We focus on the longer nanotube lengths (20 and 50 µm), for which end effects should be small, allowing us to neglect $G_c$. Here we consider the thermal model illustrated in Figure 4b. Cooling into the substrate, which was previously described by a single thermal conductance $G_{sub}$, is separated into two thermal elements in series. The first is the thermal conductance between the electron system and the phonon system in the nanotube, $G_{e-ph}$. The electron system is described by a temperature $T_e$ and a heat capacity $C_e$; the phonon system in the nanotube is described by a temperature $T_{ph}$ and a heat capacity $C_{ph}$. The phonon system in the nanotube is coupled to the substrate via the phonon interface thermal conductance $G_{int}$. Once phonons enter the substrate, they rapidly disperse into the large volume. The thermal resistances add, so $G_{sub}^{-1} = G_{e-ph}^{-1} +$ $G_{int}^{-1}$. This type of model has previously been used to describe the cooling of the electron system in metal films and long wires.[16-18]

The total thermal conductance $G$ is related to the energy relaxation time $\tau_{en}$ through the heat capacity $C$, $\tau_{en} = C/G$. This is the timescale for an excited electron, with a typical excitation energy $\sim k_B T_e$, to relax back to the Fermi energy $E_F$. The energy relaxation time due to electron-acoustic phonon scattering is $\tau_{en,e-ph} = C_e/G_{e-ph}$, where $C_e$ is the electronic heat capacity, calculated for a metallic SWNT as $C_e = 8\pi^2 L k_B^2 T_e/(3h v_F)$.[19] A different but related physical quantity is the electron-acoustic phonon scattering time $\tau_{e-ph}$. Due to momentum and energy conservation in this 1D system, the electron momentum is reversed by each scattering event.[2] $\tau_{e-ph}$ is predicted to be inversely proportional to temperature, with $\tau_{e-ph} \sim 1$ ps at room temperature.[2,9] If each acoustic phonon scattering event results in an energy change $\approx k_B T_e$, then $\tau_{en,e-ph} \approx \tau_{e-ph} \propto T^{-1}$. We believe that this is the case in our experiment, due to the use of a large gate voltage. (For further details, see the supporting information.) Hence, if $\tau_{en} \approx \tau_{e-ph}$, $G_{e-ph} \approx C_e/\tau_{e-ph} \propto T_e^2$. For $L = 50$ μm and $T_e = 100$ K, this predicts $G_{e-ph} \sim 10^{-8}$ W/K. This value is approximately consistent with the measured $G$ for $L = 50$ μm at $T_e = 100$ K, although the temperature-dependence of the measured $G$ is somewhat weaker than $T_e^2$.

We next consider the contribution of the phonon interface conductance $G_{int}$. As our experimental approach does not measure the phonon temperature, we compare our results to previous experimental determinations of $G_{int}$ at higher temperature. Shi *et al.* determine the lattice temperature in individual Joule-heated single-walled nanotubes with lengths $\approx$ 2-3 μm on $SiO_2$ at $T_b = 300$ K. In that work, they found an increase of 80-240 K at the center of the nanotube for $\approx 12.5$ μW dissipated Joule power, corresponding to $G_{int}$ $\sim 10^{-7}$-$10^{-8}$ W/K.[6] Measurements of the variation of $T_{ph}$ along the length of the nanotube established that approximately 80% of the heat loss was into the substrate ($G_{int}$), with the remainder into the contacts. Thus, for a much longer nanotube, as we have studied, cooling by phonons out the ends will be negligible.

If we extrapolate our measured $G(T_e)$ from below 120 K (below the onset of the $T_e^5$ behavior) up to $T_e \approx$ 400-500 K, we obtain values of $G$ per unit length that are approximately consistent with the experimental determination of $G_{int}$ from Ref. [6]. This extrapolation assumes that the temperature-dependence of G for small current remains essentially unchanged from 100 to 400 K, which is not yet established. Calculations of the heat flow through a nanoscale constriction at the nanotube-substrate interface predict a thermal conductance that scales linearly with the phonon heat capacity and hence with $T_{ph}$.[20] However, the interface between the nanotube and the substrate may be more complex than a nanoscale constriction, and hence the temperature dependence of $G_{int}$ may differ from the linear prediction.

Estimates of both $G_{e\text{-}ph}$ and $G_{int}$ for our temperature range yield values that are roughly consistent with our measured value of G. This suggests that both mechanisms are relevant to our measured thermal conductance. $G_{e\text{-}ph}$ is predicted to scale as $T_e^2$, and simple theory predicts that $G_{int}$ is linear in temperature. The observed temperature dependence of G at low bias current is in between these two predictions. For $L = 50$ µm, the length for which end effects should be smallest, the temperature dependence of the measured G appears to transition from approximately $T_e^2$ below 50 K to approximately $T_e$ above 50 K (Figure 3). This is consistent with $G_{sub}$ being limited by $G_{e\text{-}ph} \propto T^2$ below 50 K and by $G_{int} \propto T$ above 50 K.

We now focus on the behavior above 120 K, where the thermal conductance begins to increase rapidly with increasing temperature, approximately as $T_e^5$ (Figure 3). This behavior could be due to the bias potential between inelastic scattering events exceeding the threshold energy for optical phonon emission. The electron mean free path $l_e$ can be determined from $l_e = (h/4e^2)/(R_{int}/L)$.[2,8] At the onset of the $T_e^5$ behavior, the mean free path is approximately 2 µm. The potential energy drop over a length of 2 µm is ≈ 20 meV for all four lengths at this onset. Thermal broadening of the energy distribution is approximately $k_B T_e \approx 10$ meV at $T_e = 120$ K. The resulting energy of ≈ 30 meV is close to the 50 meV predicted threshold energy for the emission of surface polar phonons (SPPs) directly into the $SiO_2$ substrate.[21-23] Emission of optical phonons internal to the nanotube, in comparison, has a predicted threshold energy of 160-200 meV.[1,2] This 160-200 meV threshold energy is significantly larger than inferred by the data, and only the emission of surface polar phonons directly into the $SiO_2$ substrate is approximately consistent with the behavior we observe above 120 K. Shorter nanotubes have a somewhat lower onset temperature. These nanotubes have a larger thermal conductance due to cooling by the contacts. For the same electric field per unit length in the shorter nanotubes, a lower temperature is achieved. Thus, the onset of the SPP emission should occur at lower temperature for shorter nanotube lengths. We note that the approximately $T_e^5$ dependence of the thermal conductance in this regime is an empirical observation.

Calculations of dc I-V curves in the presence of strong SPP scattering found that the inverse current depends linearly on the inverse applied field,[24] $I_{dc}^{-1} = I_s^{-1} + R_{0,int}V_{int}^{-1}$, where $I_s$ is the saturation current due to SPP scattering, $R_{0,int}$ is the internal nanotube resistance near zero bias current, and $V_{int}$ is the voltage drop internal to the nanotube (excluding the contacts). Our measured dc I-V curves agree well with this functional form for $I_{dc} > 3.5$ µA. Fitting to I-V curves measured at $T_b = 77$ K, with $I_s$ as a fitting parameter, we find $I_s \approx 12$ µA for all four nanotube lengths. By comparison, Ref. [24] calculates $I_s = 4.4$ µA for a (17,0) semiconducting nanotube on a quartz substrate at $T_b = 77$ K. In this calculation, the doping level is 0.1 e/nm and the tube-substrate separation is 3.5 Å.[23] The difference in $I_s$ is likely the result of the differences between this modeled system and our measured sample. In particular, $I_s$ is strongly dependent on the tube-substrate separation. We note that the effects of SPPs have previously been observed in the electrical transport properties of individual semiconducting SWNTs on $SiO_2$ at $T_b = 300$ K,[25] but have not previously been observed in a metallic nanotube or at substrate temperatures below room temperature.

Finally, we propose an experiment for further elucidating the limiting cooling mechanism for the longer nanotube lengths in low-bias regime ($T_e < 120$ K). The energy relaxation time $\tau_{en}$ could be measured directly by heating the nanotube with a fast voltage pulse and measuring the exponential decay of the resistance back to its equilibrium value. Alternatively, $\tau_{en}$ could be measured in the frequency domain via heterodyne mixing.[16] If the limiting cooling mechanism is acoustic phonon emission, then the relevant heat capacity is the electronic heat capacity $C_e = 8\pi^2 L k_B^2 T_e/(3hv_F)$,[19] and $\tau_{en} = \tau_{en,e-ph}$. For $L = 50$ µm and $T_e = 100$ K, $\tau_{en,e-ph} = C_e/G \approx 6$ ps. If instead the cooling is limited by heat escaping the phonon system, then the relevant heat capacity is the phonon heat capacity $C_{ph} \approx C_e(v_{ph}/v_F)$,[19] where the ratio of the acoustic phonon velocity to the Fermi velocity $v_{ph}/v_F \approx 1/50$. Hence the time constant will be $\tau_{en} = C_{ph}/G \approx 300$ ps. Measuring this slower time constant is feasible,[16] but it requires a sample that, unlike the sample studied in the present work, is designed without significant parasitic reactance at microwave frequencies.

In addition to a direct measurement of the energy relaxation time, future measurements using these techniques should further refine our understanding of the energy loss mechanisms of the non-equilibrium nanotube electron system. Promising avenues for further study include measurements of suspended nanotube samples and measurements of the dependence of the thermal conductance on gate voltage. The techniques described here may also prove useful in similar studies of inelastic scattering in other conducting nanosystems.

**Supporting Information Available**

A discussion of the connection between the electron-phonon scattering time and the energy relaxation time, as well as an experimental test of the thermal conductance.

**Acknowledgement**

We thank A. J. Annunziata, M. Brink, L. Frunzio, A. Kamal, P. L. McEuen, B. Reulet, and S. V. Rotkin for helpful discussions. The work at Yale was supported by NSF grants DMR-0907082 and CHE-0911593. P. K. and M. S. P. acknowledge financial support from NSF NIRT (ECCS-0707748) and Honda R&D Co.

**Supporting Information**

"Energy loss of the electron system in individual single-walled carbon nanotubes,"

Daniel F. Santavicca, Joel D. Chudow, Daniel E. Prober, Meniner S. Purewal, and Philip Kim

## I. Test of experimentally-determined thermal conductance

In the absence of non-thermal nonlinearities, the differential resistance $dV/dI$ is related to the dc resistance $R_{dc} = V_{dc}/I_{dc}$ through the thermal conductance $G$,[1]

$$\frac{dV}{dI} = R_{dc} \frac{1 + \frac{P}{R_{dc} G} \frac{dR}{dT}}{1 - \frac{P}{R_{dc} G} \frac{dR}{dT}} \tag{S1}$$

where $P$ is the dc Joule power, in this case the power dissipated internal to the nanotube, $P_{NT}$. In Figure S1 we plot for the 5 μm nanotube at $T_b = 4.2$ K the differential resistance calculated from eq S1 using the experimentally-determined $G$ as well as measured values of $R_{dc}$, $P_{NT}$, and $dR/dT$. $dV/dI$ is only calculated for $T_e > 20$ K because of the ZBA feature at lower temperature. In this calculation, we assume that only $R_{int}$ is heated and that $R_c$ is a temperature-independent resistor. We also plot the measured $dV/dI$ and $R_{dc}$ as a function of $I_{dc}$. The good agreement between the measured and calculated values of $dV/dI$, with no adjustable parameters, supports our experimental determination of the thermal conductance.

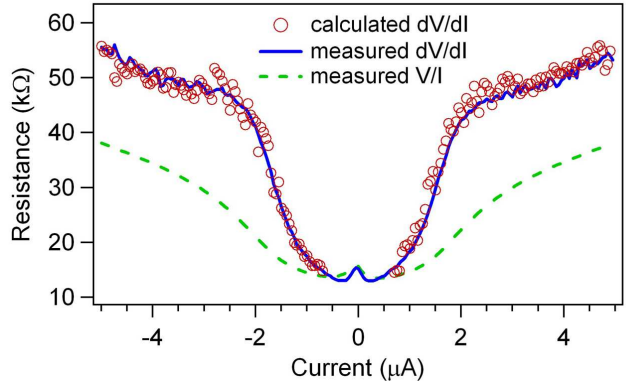

FIGURE S1. Measured and calculated differential resistance, as well as measured dc resistance, for 5 μm nanotube at $T_b = 4.2$ K.

## II. Comparison of energy relaxation and electron-phonon scattering times

The energy relaxation time $\tau_{en}$ is the characteristic time for an excited electron, with a typical excitation energy $\sim k_B T_e$, to relax back to the Fermi level $E_F$. The electron-acoustic phonon time $\tau_{e\text{-}ph}$ is the characteristic timescale for an electron scattering event due to acoustic phonons. Due to momentum and energy conservation in the 1D nanotube, the electron momentum is reversed by each scattering event, as illustrated in Figure S2. The electron energy change per acoustic phonon scattering event depends on the Fermi level, which depends on the gate voltage. Due to the linear electron and phonon dispersion relations at low energies, the energy of an emitted acoustic phonon is $E_{ph} \approx 2E_{el}(v_{ph}/v_F)$.[2] The electron energy $E_{el}$ has a range of approximately $E_F \pm k_B T_e$, where $E_F$ is defined relative to the band crossing. The ratio of the phonon velocity to the Fermi velocity $(v_{ph}/v_F) \approx 1/50$.[2] Thus, for $E_F = 0$, $E_{ph} \sim k_B T_e/25$, and many scattering events are required to remove an energy of $k_B T_e$. This is illustrated schematically in Figure S2a. As $E_F$ is increased by applying a gate voltage, $E_{ph}$ will increase, as illustrated in Figure S2b. In the present work, we use a large gate voltage, -30 V. Accounting for gate hysteresis and assuming a gate efficiency $\sim 1\%$,[3] we estimate $E_F \sim 0.1$ eV. This yields $E_{ph} \sim k_B T_e$ in our experimental range. Hence, in the present work, $\tau_{e\text{-}ph}$ is believed to be approximately equal to the energy relaxation time $\tau_{en}$.

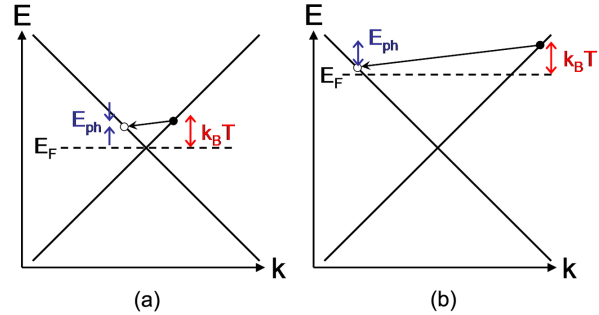

FIGURE S2. (a) Electron dispersion relation illustrating the emission of an acoustic phonon at zero gate voltage. (c) Electron dispersion relation illustrating the emission of an acoustic phonon at finite gate voltage.

## References
[1] M. Galeazzi and D. McCammon, *J. Appl. Phys.* **2003**, *93*, 4856.
[2] J.-Y. Park, S. Rosenblatt, Y. Yaish, V. Sazonova, H. Usunel, S. Braig, T. A. Aria, P. W. Bouwer and P. L. McEuen, *Nano Lett.* **2004**, *4*, 517-520.
[3] J. Cao, Q. Wang and H. Dai, *Nat. Mater.* **2005**, *4*, 745.

16